\documentclass[prl,twocolumn,nofootinbib,showpacs,superscriptaddress]{revtex4}

\usepackage[english]{babel}
\usepackage{amsmath,amssymb}
\usepackage[dvips]{graphicx}
\usepackage{ifthen,array}
\usepackage[sort&compress]{natbib}

\newcommand{\pref}[1]{(\ref{#1})}

\newcommand{\AHEP}{Instituto de F\'{\i}sica Corpuscular --
  C.S.I.C./Universitat de Val{\`e}ncia \\
  Edificio Institutos de Paterna, Apt 22085,
  E--46071 Val{\`e}ncia, Spain}

\newcommand{\IZMIRAN}{The Institute of Terrestrial Magnetism,
  Ionosphere and 
  Radio Wave Propagation \\ Russian Academy of Sciences,
  IZMIRAN, Troitsk, Moscow region, 142190, Russia}

\begin{document}

\vspace*{3mm}
\title{Cornering Solar Radiative-Zone Fluctuations with KamLAND
  and SNO Salt}

\author{C.P. Burgess}
\affiliation{Physics Department, McGill University,
  3600 University Street, Montr\'eal, Qu\'ebec, Canada H3A 2T8.}

\author{N.S. Dzhalilov}
\affiliation{\IZMIRAN}

\author{M. Maltoni}
\affiliation{\AHEP}

\author{T.I. Rashba}
\affiliation{\IZMIRAN}
\affiliation{\AHEP}

\author{V.B. Semikoz}
\affiliation{\IZMIRAN}
\affiliation{\AHEP}

\author{M.A. T\'ortola}
\affiliation{\AHEP}

\author{J.W.F. Valle}
\affiliation{\AHEP}

\begin{abstract}
    We update the best constraints on fluctuations in the solar medium
    deep within the solar Radiative Zone to include the new SNO-salt
    solar neutrino measurements. We find that these new measurements
    are now sufficiently precise that neutrino oscillation parameters
    can be inferred independently of any assumptions about fluctuation
    properties. Constraints on fluctuations are also improved, with
    amplitudes of 5\% now excluded at the 99\% confidence level for
    correlation lengths in the range of several hundred km. Because
    they are sensitive to correlation lengths which are so short,
    these solar neutrino results are complementary to constraints
    coming from helioseismology. 
\end{abstract}


\maketitle

\section{Introduction}

Neutrino-oscillation measurements are entering an era of unprecedented
precision, with the solar neutrino data
\cite{sol02,Ahmad:2002jz,Fukuda:2002pe} and atmospheric neutrino data
\cite{atm02,Fukuda:1998mi} combining to give a concordant picture of
conversions amongst three species of active
neutrinos~\cite{Maltoni:2003da,pakvasa:2003zv}.  The oscillation
parameters which describe these conversions are the two mass-squared
differences, $\Delta m^2_{\mathrm{sol}}$ and $\Delta
m^2_{\mathrm{atm}}$, the three mixing angles, $\theta_{12}$,
$\theta_{23}$ and $\theta_{13}$, plus phases which violate CP
\cite{Schechter:1980gr} and are still to be probed. The best fits to
these parameters are consistent with a maximal atmospheric mixing
angle, $\theta_{23}$, and give a preferred solar mixing angle in the
so-called large mixing angle (LMA-MSW) regime
\cite{Gonzalez-Garcia:2000aj}. The third angle, $\theta_{13}$, is
strongly constrained mainly by reactor experiments
\cite{Apollonio:1999ae}.

A crucial recent development has been the verification of these
oscillation parameters in purely terrestrial measurements, with the
KamLAND experiment \cite{kamland} reporting measurements which are
consistent with the oscillation parameters indicated by the solar
neutrino analysis. Such an independent measurement of oscillation
properties is invaluable since it allows a cleaner separation to be
made between neutrino properties and solar physics, thereby opening a
new observational window deep into the solar interior \cite{Bahcall}.

In particular, the precise terrestrial observation of oscillations
relevant to solar neutrinos allows the removal of a theoretical
uncertainty in the inference of neutrino properties. This uncertainty
arises because specific types of fluctuations of the solar medium deep
within the solar radiative zone are known to affect neutrino
oscillations
\cite{Balantekin:1996pp,Nunokawa:1996qu,Burgess:1996,Bamert:1997jj},
if they have sufficient size. Consequently the inference of neutrino
properties from solar data require the use of prior assumptions
concerning these such fluctuations.

Traditionally, the necessity for making these prior assumptions
concerning solar fluctuations has not been regarded as being
worrisome for several reasons. First, helioseismic measurements
can constrain deviations of solar properties from Standard Solar
Model predictions at better than the percent level. Second,
preliminary studies of the implications for neutrino oscillations
of radiative-zone helioseismic waves \cite{Bamert:1997jj} showed
that they were very unlikely to have observable effects. Third, no
other known sources of fluctuations seemed to have the properties
required to influence neutrino oscillations.

All three of these points have been re-examined in recent years, with
the result that the presence of solar fluctuations seems more likely
than previously thought. First, direct helioseismic bounds turn out to
be insensitive to fluctuations whose size is as small as those to
which neutrinos are sensitive
\cite{Castellani:1997pk,Christensen-Dalsgaard:2002ur} (which, as we
argue below, turn out to be those whose size is only several hundreds
of km). Second, recent studies of magnetic fields deep inside the
solar radiative zone \cite{heliomag} have identified potential
fluctuations to which neutrinos might be sensitive after all (due to a
resonance between Alfv\'en waves and helioseismic $g$-modes).

These studies motivate us to investigate again the extent to which
neutrino-oscillation parameters can be extracted independent of prior
assumptions concerning the solar fluctuations. In principle,
sufficiently precise separate measurement of oscillation parameters by
KamLAND and by solar neutrino detectors may allow this type of prior
assumption to be relaxed. Unfortunately, global fits to the
post-KamLAND data in the presence of fluctuations~\cite{us,them} have
indicated that the data were not yet sufficiently precise to allow the
neutrino-oscillation parameters and the solar fluctuations to be
disentangled with good accuracy.

The recent release of the SNO salt results \cite{SNOsalt} call for a
re-evaluation of this conclusion, since these considerably improve the
precision with which solar neutrino properties are determined. It is
the purpose of the letter to show that with these new data solar
neutrino measurements have now changed the picture, inasmuch as global
fits to neutrino properties no longer need to make prior assumptions
about solar medium fluctuations in order to infer neutrino oscillation
parameters. We also summarize the direct constraints on these
fluctuations which may now be convincingly inferred for the first
time, without making prior assumptions concerning the details of
neutrino oscillations.  Finally we forecast the precision which will
be possible to achieve with subsequent terrestrial neutrino
measurements.


\section{Sensitivity to Fluctuations}

The standard description of MSW oscillations
\cite{Wolfenstein:1977ue} amount to the use of a mean-field
approximation for the solar medium. The corrections to this
mean-field approximation are due to the fluctuations in the solar
medium about this mean, and the leading interaction of neutrinos
with these fluctuations are parameterized by the electron-density
autocorrelation, $\langle \delta n_e(t) \delta n_e(t') \rangle$,
measured along the neutrino trajectory
\cite{Balantekin:1996pp,Nunokawa:1996qu,Burgess:1996,Bamert:1997jj}.

\begin{figure}[t] \centering
    \includegraphics[width=0.99\linewidth]{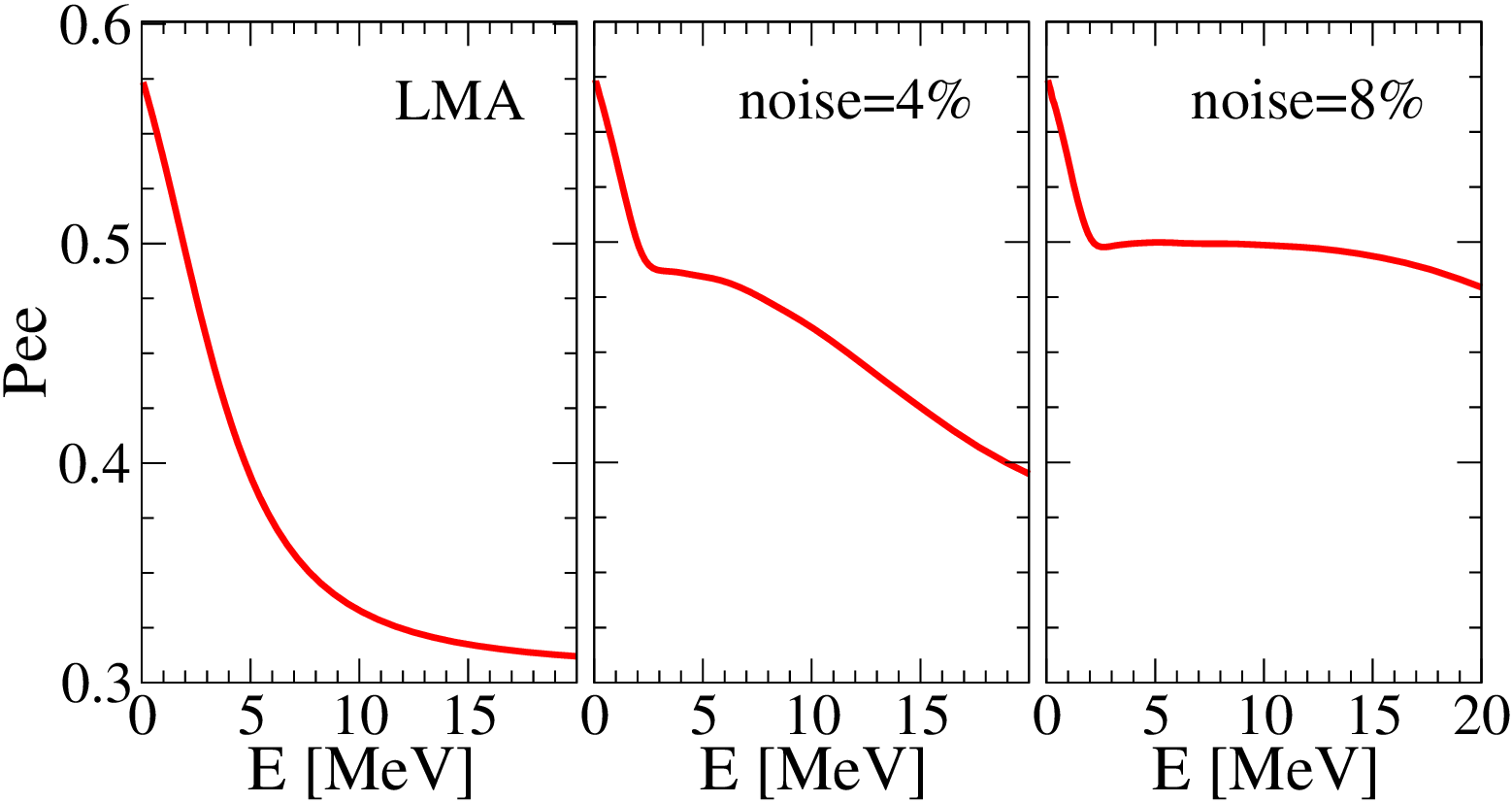}
    \caption{\label{fig:noisyLMA}%
      Effect of random electron density fluctuations on
      electron-neutrino survival probability for LMA-MSW oscillations
      and a correlation length $L_0 = 100$ km.  The fluctuation's
      amplitude $\xi$ at the position of neutrino resonance is zero in
      the left panel, and is $\xi=4 \%$ and $\xi=8 \%$ in the middle
      and right panels, respectively.}
\end{figure}

As fig.~\pref{fig:noisyLMA} shows, such fluctuations act to degrade
the efficiency of neutrino conversions. They can do so because
successive neutrinos `see' slightly different solar properties, and so
in particular do not experience an equally adiabatic transition as
they pass through the neutrino resonance region. The net effect is to
degrade the effectiveness of the neutrino conversion because those
neutrinos for which the transition is less adiabatic are more likely
to survive as electron-type neutrinos. Since criterion for the
transition to be adiabatic depends on how quickly the electron
distribution varies near resonance, fluctuations give observable
effects for neutrinos if they occur at resonance with sufficient
amplitude, and if their correlation length, $L_0$, is comparable to
the local neutrino oscillation length, $L_{\rm osc} \sim 100$ km.

\section{The Implications of the SNO Salt Result}

We now report on the result of fits which are obtained using a global
analysis of the solar data, including radiochemical experiments
(Chlorine, Gallex-GNO and SAGE)~\cite{sol02} as well as the latest SNO
data in the form of 17 (day) + 17 (night) recoil energy bins (which
include CC, ES and NC contributions,
see~\cite{Maltoni:2002ni})~\cite{Ahmad:2002jz} and the
Super-Kamiokande spectra in the form of 44 bins~\cite{Fukuda:2002pe}
(8 energy bins, 6 of which are further divided into 7 zenith angle
bins). We have also used the improved measurement with enhanced
neutral current sensitivity due to neutron capture on salt, which has
been added to the heavy water in the SNO detector.  The data are
presented in the form of the neutral current (NC), charged current
(CC) and elastic scattering (ES) fluxes~\cite{SNOsalt}. Following
Ref.~\cite{Maltoni:2002aw} we used data from the KamLAND collaboration
given in 13 bins of prompt energy above 2.6 MeV~\cite{kamland}.

\begin{figure}[t] \centering
    \includegraphics[width=0.99\linewidth,]{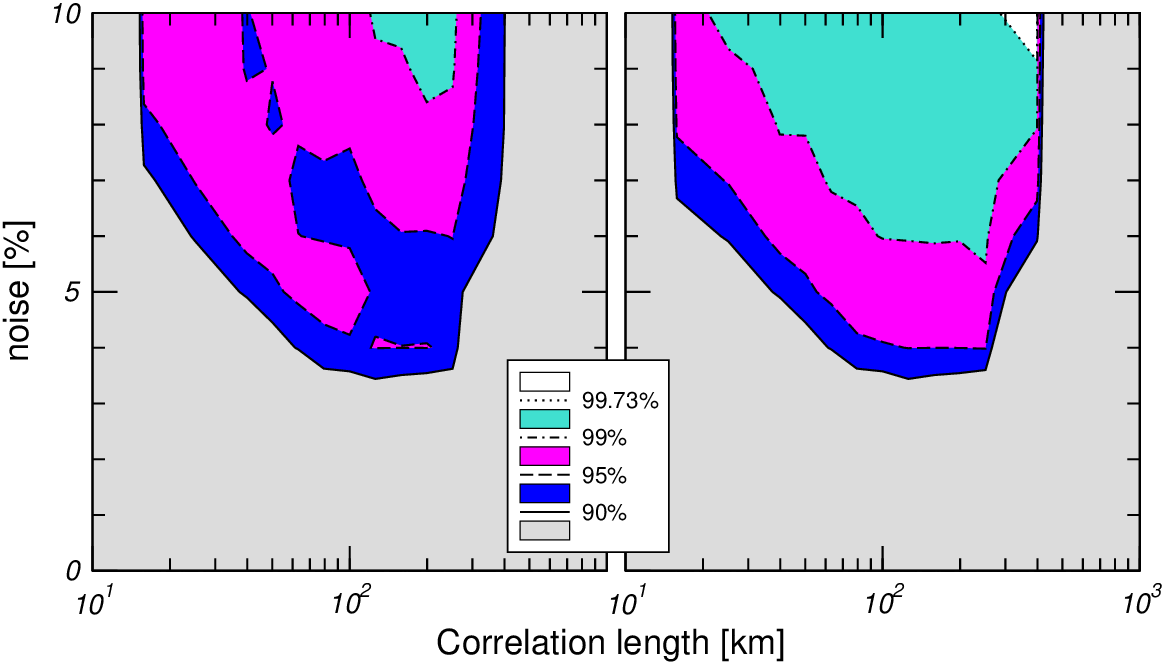}
    \caption{\label{fig:chi2fit}%
      Exclusion region in the amplitude--correlation length ($\xi -
      L_0$) plane for solar fluctuations using KamLAND and solar
      neutrino data before the SNO--salt experiment. In the right
      panel the neutrino oscillation parameters are assumed known
      while both oscillation and fluctuation parameters are jointly
      fit in the left panel. The lines indicate contours of 90, 95,
      99\% CL and 3~$\sigma$.}
\end{figure}

\begin{figure}[t] \centering
    \includegraphics[width=0.99\linewidth]{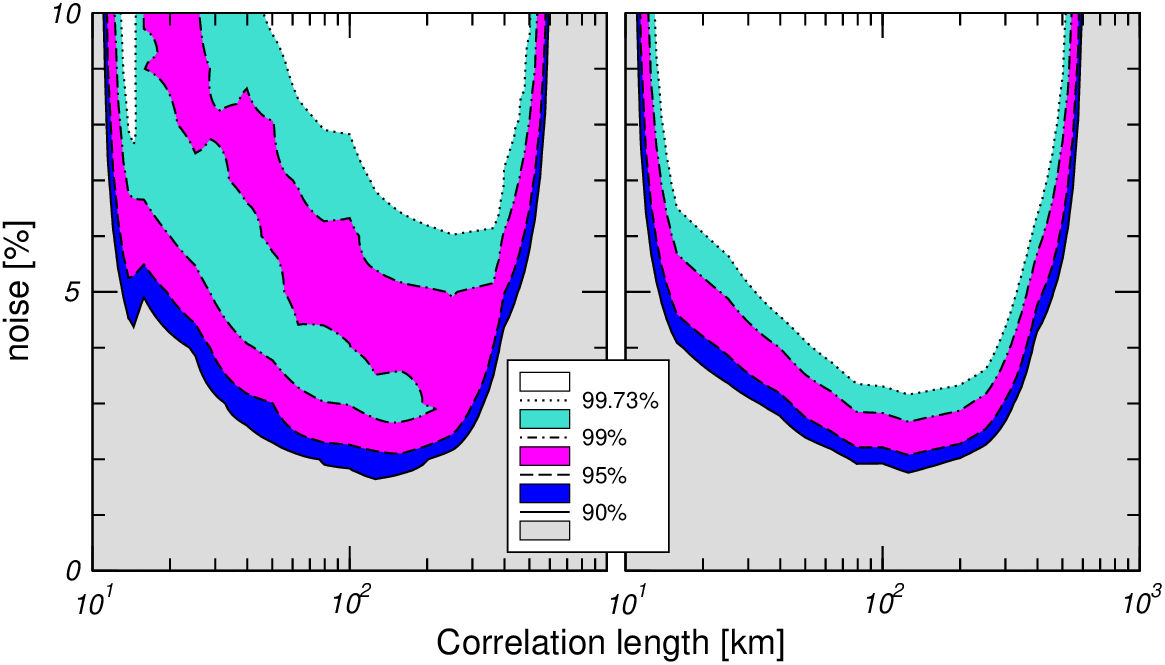}
    \caption{\label{fig:noisyLMAsalt}%
      Sensitivity of solar neutrino data to the solar fluctuations
      including the SNO salt measurements. As in
      Fig.~\ref{fig:chi2fit} the right panel assumes the neutrino
      oscillation parameters are known while the left panel shows the
      result when both oscillation parameters and fluctuations are
      jointly fit.}
\end{figure}

The sensitivity of the solar neutrino data to fluctuations in the
solar medium is summarized by figures \pref{fig:chi2fit} and
\pref{fig:noisyLMAsalt}. Fig.~\pref{fig:chi2fit} is taken from
ref.~\cite{us}, and summarizes the sensitivity before the SNO salt
measurements. Fig.~\pref{fig:noisyLMAsalt} gives the same results
after SNO salt. Comparing these figures shows the improvement in
constraints due to the SNO salt data, and comparing the panels in
each figure shows the importance of a precise determination of
the neutrino oscillation parameters for obtaining a constraint on
the magnitude of fluctuations.

\begin{figure}[t] \centering
    \includegraphics[width=0.99\linewidth]{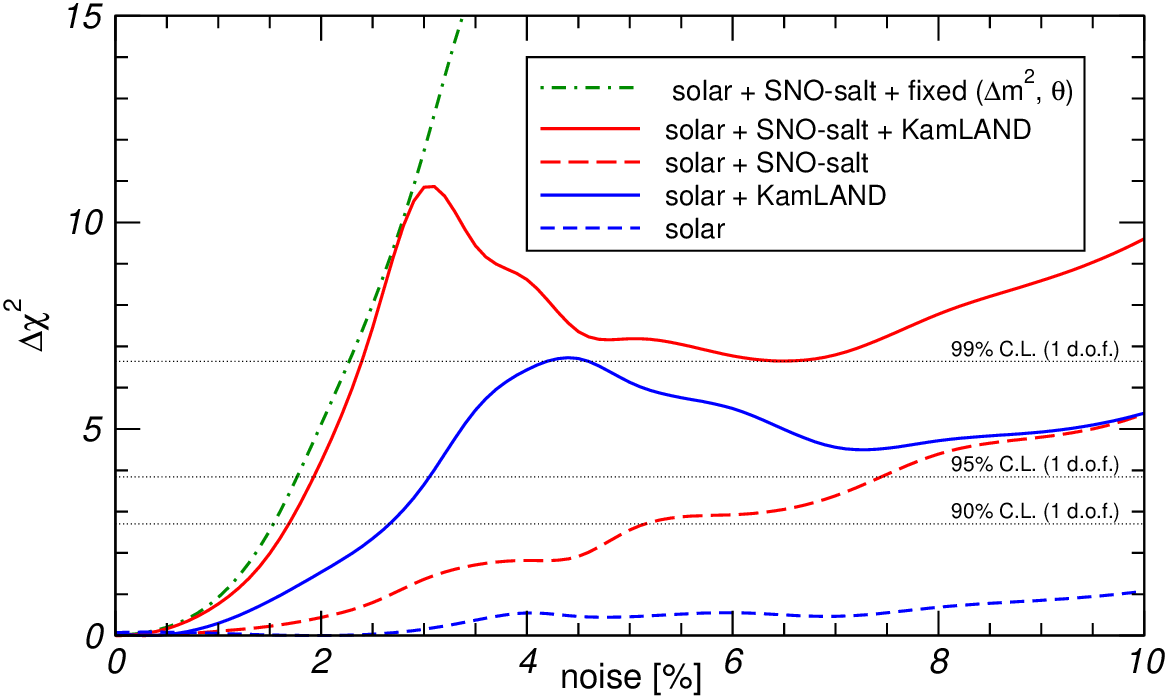}
    \caption{\label{fig:noisyLMAchisq}%
      The chi-square of the fit as a function of fluctuation amplitude
      for a correlation length $L_0 = 100$ km.}
\end{figure}

The importance of the KamLAND and the SNO-salt measurements in these
results is most easily seen from fig.~\pref{fig:noisyLMAchisq}, which
compares the dependence of the fit's $\chi^2$ on the amplitude of the
fluctuations for various individual data sets. This figure makes clear
how the KamLAND results are largely responsible for localizing the
best fit near zero fluctuation amplitude.  This is as should be
expected, since the evidence for the absence of fluctuations follows
from the comparison of solar neutrino observations with terrestrial
measurements of neutrino oscillation properties. 

Note, however, that further precise determination of neutrino
parameters at KamLAND due to higher statistics will have a very modest
impact on the limit on the amplitude of density fluctuations within
the 99\% C.L. region, as can be seen from the dot--dashed line in
Fig.~\ref{fig:noisyLMAchisq}.

\begin{figure}[t] \centering
    \includegraphics[width=0.99\linewidth]{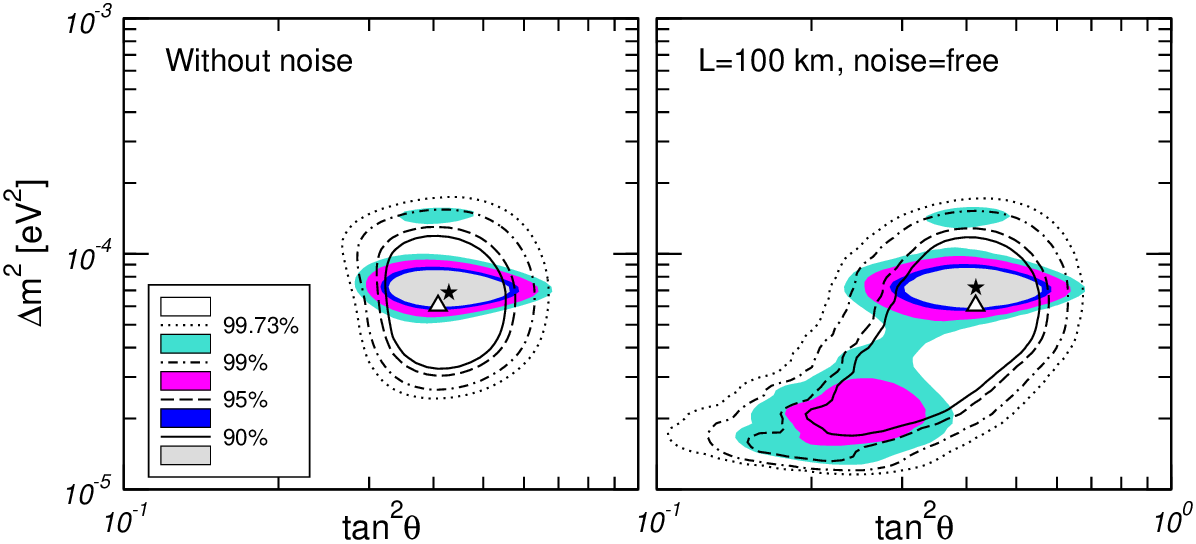}
    \caption{\label{fig:LMAwwonoise}%
      The solar neutrino oscillation parameters obtained in our 
      global fit for $L_0 = 100$ km.  The left panel assumes no noise,
      while in the right panel the amplitude of the noise is left
      arbitrary. The coloured regions are obtained using the KamLAND
      data, while the lines refer to CL contours without KamLAND.}
\end{figure}

Fig.~\pref{fig:LMAwwonoise} shows how the existence of solar
fluctuations influences the determination of the neutrino oscillation
parameters, and contains our main result. The two panels of the figure
contrast the precision of the fit with and without solar fluctuations.
The left panel gives results subject to the usual prior assumption of
no solar fluctuations, while the right panel leaves the amplitude of
such fluctuations to be obtained from the fit. (When fluctuations are
included, they are assumed to have the optimal correlation length $L_0
= 100$ km.)  The lines indicate contours of fixed confidence level
when the KamLAND data are not included, while the coloured regions
give the same information when KamLAND is included.

\begin{figure}[t] \centering
    \includegraphics[width=0.99\linewidth]{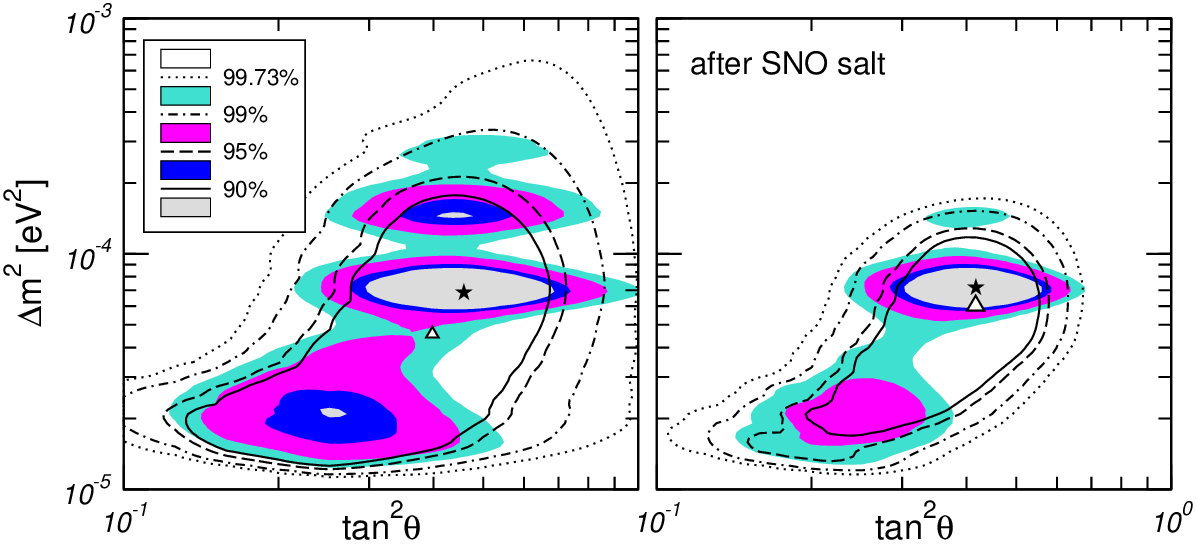}
    \caption{\label{fig:noisyLMAbeforesalt}%
      The same fit (including fluctuations) as given in
      fig.~\pref{fig:LMAwwonoise}, performed with (right) and without
      (left) the SNO salt results.}
\end{figure}

The main conclusion which follows from this figure is that the
precision with which the neutrino oscillations are known is now
largely independent of whether a prior assumption is made about
the existence of solar fluctuations. With the release of the SNO
salt results the comparison of solar neutrino with KamLAND data
suffices to robustly determine the oscillation parameters
independent of the assumed amplitude of solar fluctuations. The
SNO salt data are crucial for reaching this conclusion, as is
clear from fig.~\pref{fig:noisyLMAbeforesalt}, which compares the
right panel of fig.~\pref{fig:LMAwwonoise} with the same fit
performed without using the SNO salt results.

\section{Outlook}

We see that the SNO salt data, when combined with KamLAND results, for
the first time places the determination of neutrino-oscillation
parameters beyond the reach of sensitivity to prior assumptions
concerning the existence of fluctuations in the solar radiative zone.
Besides making more robust the determination of neutrino-oscillation
parameters, this allows a much sharper determination of the kinds of
solar fluctuations that may still be allowed deep within the solar
radiative zone. As we have seen, the resulting constraints apply to
fluctuations whose spatial scales are of order 100 km, and so are
complementary to those obtained from helioseismology, which are
insensitive to fluctuations on such short scales.

Ref.~\cite{heliomag} has suggested one possible mechanism for
obtaining observable fluctuations in the relevant part of the sun.  In
that picture fluctuations having the appropriate distance scale may
arise if magnetic fields of order 10 kG should exist deep in the solar
core. Magnetic fields of this size would be consistent with current
observational bounds \cite{Couvidat:2002bs,Friedland}.

It is instructive to ask how the precision of these results is likely
to improve given the new neutrino experiments which are currently
being planned. In addition to further statistics from the KamLAND
reactor experiment, we expect a high statistics of solar neutrinos
above 5 MeV at the UNO experiment, which would make solar precision
measurements possible~\cite{Yanagisawa}.  Quantifying the sensitivity
of upcoming detectors as probes of the Sun deep within the radiative
zone lies outside the scope of this note.

\vspace*{3mm}

Work supported by Spanish grants BFM2002-00345, by the European
Commission RTN network HPRN-CT-2000-00148, by the European Science
Foundation network grant N.~86, by CSIC-RAS agreement (VBS) and by
MECD grants SB-2000-0464 (TIR) and AP2000-1953 (MAT). C.B.'s research
is supported by grants from NSERC (Canada), FCAR (Quebec) and McGill
University. VBS, NSD and TIR were partially supported by the program
of Presidium RAS ``Non-stationary phenomena in astronomy''.

\end{document}